\documentclass[12pt]{article}
\usepackage{amsmath}
\usepackage[dvips]{graphicx}

\newcommand{\bm}{\begin{multiline}}
\newcommand{\beq}{\begin{equation}}
\newcommand{\eeq}{\end{equation}}
\newcommand{\beqs}{\begin{eqnarray}}
\newcommand{\eeqs}{\end{eqnarray}}
\newcommand{\beqa}{\begin{eqnarray}}
\newcommand{\eeqa}{\end{eqnarray}}
\newcommand{\beqar}{\begin{eqnarray*}}
\newcommand{\eeqar}{\end{eqnarray*}}

\begin{document}

\addtolength{\baselineskip}{1.5mm}

\thispagestyle{empty}

\hfill{} \vspace{32pt}

\begin{center}
\textbf{Generation of Cosmological Seed Magnetic Fields from Inflation with
Cutoff}\\[0pt]

\vspace{48pt} Amjad Ashoorioon\footnote{%
amjad@astro.uwaterloo.ca}, Robert B. Mann\footnote{%
mann@avatar.uwaterloo.ca}

\vspace{12pt} \textit{Department of Physics, University of Waterloo,
Waterloo, Ontario, N2L 3G1, Canada}
\end{center}

Inflation has the potential to seed the galactic magnetic fields observed
today. However, there is an obstacle to the amplification of the quantum
fluctuations of the electromagnetic field during inflation: namely the
conformal invariance of electromagnetic theory on a conformally flat
underlying geometry. As the existence of a preferred minimal length breaks
the conformal invariance of the background geometry, it is plausible that
this effect could induce electromagnetic field amplification. We show that
this scenario is equivalent to endowing the photon with a large negative
mass during inflation. This effective mass is negligibly small in a
radiation and matter dominated universe. Depending on the value of the free
parameter in the theory, we show that the seed required by the dynamo
mechanism can be generated. We also show that this mechanism can produce the
requisite galactic magnetic field without resorting to a dynamo mechanism.

\setcounter{footnote}{0} \newpage

\section{Introduction}

Cosmic magnetic fields are ubiquitous at all large intragalactic scales. It
is a well-known observational fact that our galaxy and many other spiral
galaxies are endowed with coherent magnetic fields of $\mu $G (microgauss)
strength \cite{Sofu, Asseo, Kronberg, Beck, Grasso}, having approximately
the same energy density as the cosmic microwave background radiation (CMBR).
There is also evidence for larger magnetic fields of similar strength within
clusters \cite{Perely, Kim}. The presence of magnetic fields at larger
scales has also been confirmed \cite{Giovannini,Vallee}. These magnetic
fields play an important role in various astrophysical processes, such as
the confinement of cosmic rays and the transfer of angular momentum away
from protostellar clouds so that they can collapse and become stars.
Magnetic fields are also present in the intracluster gas of rich clusters of
galaxies, in quasistellar objects (QSO's) and in active galactic nuclei.
They may influence the formation process of large-scale structure \cite%
{Subramanian,Tsagas}.

It is widely believed that galactic magnetic fields are amplified and
sustained by a dynamo mechanism \cite%
{Kronberg,Beck,Grasso,Parker1,Parker2,Zel'dovich}, in which the cyclonic
turbulent motion of ionized gas combined with the differential rotation of
the galaxy exponentially amplifies a ``seed'' magnetic field. This continues
until the backreaction of the motion of the plasma offsets the growth of the
field, stabilizing it to dynamical equipartition strength. However, while
the dynamo mechanism provides an amplification mechanism, it does not
explain the origin of galactic magnetic fields, and requires a ``coherent''
seed magnetic field for it to be effective. Indeed, it has been shown that
seed magnetic fields that are too incoherent may undermine the action of the
dynamo \cite{Kulsrud}. Most dynamo scenarios require a minimum coherence
length equal to the dimension of the largest turbulent eddy, usually around $%
\sim 100$ pc. If the mechanism has functioned over the whole age of the
galaxy($\sim $ 10G yr) a seed field of $10^{-19}$G is required. If recent
observations are correct and the universe is dominated by a dark-energy
density component \cite{Supernova, Jaffe}, then galaxies are older than
previously thought and the seed magnetic field may be as low as $%
B_{seed}\sim {10}^{-30}$G \cite{Davis}.

A contrasting view is that the primeval magnetic flux trapped in the gas
that collapsed to form the galaxy is responsible for the existence of
galactic magnetic fields. This hypothesis also requires the existence of a
seed magnetic field, one that is as great as the field observed today \cite%
{Piddington, Ohki}. Several scenarios have been suggested for creation of
the required seed magnetic field, the most important of which involve
battery \cite{Kulsrud} or vorticity \cite{Harrison} effects. The battery
mechanism requires a large-scale misalignment of density and pressure
gradients usually related to active galactic nuclei (AGN) or starburst
activity. Therefore, it is difficult to realize in the majority of galaxies.
The vorticity mechanism is based on the relative motions of photons and
electrons induced by vorticity that was present before decoupling. Of course
this mechanism assumes the existence of primeval vorticity. In addition,
large-scale vortical motions can be effective only if ionization of the
plasma is considerable, which does not occur at the galaxy-formation epoch.

Throughout most of the history of the universe the average time $\tau $%
\thinspace between particle interactions has been much smaller than the
expansion time scale, $\tau \ll t_{Hubble}$ . Consequently the universe has
been a good conductor \cite{Turner1}, and any primeval cosmic magnetic field
would have evolved in a manner that preserved magnetic flux: $Ba^{2}\sim $
constant, where $a$ is the scale factor. Hence the dimensionless ratio $%
r=B^{2}/\left( 8\pi \rho _{\gamma }\right) $ is almost constant and provides
a convenient measure of magnetic field strength. If there had been a
pregalactic cosmic magnetic field that collapsed with the gas that formed
the galaxy, its strength must have increased as $\mathbf{[\rho }_{gal}%
\mathbf{/\rho }_{tot}(t)]^{2/3}$, where $\rho _{tot}(t)$ is the average
cosmic mass density at time $t$. As $\rho _{tot}\propto a^{-3}$ and $\rho
_{gal}/\rho _{tot}=10^{6}$ today ($t_{0}=0$), it follows that the strength
of the magnetic field at the time of formation $t_{form.}$ must have been $%
10^{4}[a(t_{form.})/a(t_{0})]^{2}B_{cosmic}$ or $B_{gal.}\simeq
3r^{1/2}10^{-2}$G. This yields $r\simeq 10^{-34}$ for initiating the
galactic dynamo, or alternatively $r\simeq 10^{-8}$ for seeding the galactic
magnetic field itself while avoiding the necessity of a galactic dynamo. If
the existence of dark energy in the universe is confirmed, the minimum $r$
required to seed the dynamo mechanism reduces to $10^{-56}$.

Inflation offers the hope of furnishing a mechanism for kinematically and
dynamically producing the seed for cosmic magnetic fields. It provides the
kinematic means for producing long-wave-length effects at very early times
through microphysical processes operating on scales less than the Hubble
radius. Since an electromagnetic wave with $\lambda _{phys}\geq H^{-1}$ has
the appearance of static $\mathbf{E}$ and $\mathbf{B}$ fields, very long
wavelength photons ($\lambda _{phys}\gg H^{-1}$) can lead to large-scale
magnetic fields (which then become supported by currents). Of course the
electric field generated during the inflationary stage not only is not
amplified, but is actually damped down due to the large conductivity of the
primeval plasma. Another reason inflation is considered to be a prime
candidate for field amplification is the fact that during inflation the
universe is devoid of charged particles. Hence, the magnetic flux is not
necessarily conserved and $r$ can increase. Furthermore, inflation can
superadiabatically amplify the energy density($\simeq kd\rho /dk$) of the
minimally coupled field \cite{Turner2}. Then the energy density decays as $%
a^{-2}$, rather than the usual result $a^{-4}$(``adiabatic result'').

However the conformal flatness of the Robertston-Walker metric prevents the
background gravitational field from producing particles, provided the
underlying matter theory is conformally invariant. A pure U(1) gauge theory
with the standard lagrangian $\mathcal{L}=-\frac{1}{4}F_{\mu \nu }F^{\mu \nu
}$ is conformally invariant, from which it follows that $\rho _{B}\propto
B^{2}$ always decreases as $a^{-4}$. During the inflationary epoch, the
total energy density in the universe is dominated by vacuum energy and
therefore the energy density in any magnetic field during inflation is
significantly suppressed. In fact it can be shown that $r=10^{-104}\lambda
_{Mpc}$ \cite{Turner1}, where $\lambda_{Mpc}\equiv \lambda/1 Mpc$.

Several proposals have been given to break the conformal invariance of the
theory: (i) coupling the electromagnetic field to a
non-conformally-covariant charged field \cite{Ratra}, (ii) coupling the
electromagnetic field to gravity via either gauge non-invariant terms such
as $RA_{\mu }A^{\mu }$, $R_{\mu \nu }A^{\mu }A^{\nu }$ or gauge invariant
ones like $R_{\mu \nu \lambda \kappa }F^{\mu \nu }F^{\lambda \kappa }/m^{2}$%
, $R_{\mu \nu }F^{\mu \kappa }F_{\kappa }{}^{\nu }/m^{2}$ or $RF^{\mu \nu
}F_{\mu \nu }/m^{2}$ \cite{Turner1}, (iii) invoking effects due to the
quantum conformal anomaly \cite{Dolgov,Chanowitz}, (iv) creating primordial
magnetic fields at either the QCD transition epoch \cite{Quashnock} or the
electroweak transition \cite{Vachaspati}, (v) breaking conformal invariance
via nonzero vacuum expectation values of flat directions in minimally
supersymmetric standard models (MSSM) \cite{Enqvist}.

Attempts to realize the first possibility were carried out by coupling the
electromagnetic field to the scalar field $\Phi $ responsible for inflation
via a term $\propto e^{\Phi }F^{\mu \nu }F_{\mu \nu }$ \cite{Ratra}. This
investigation showed that in the exponential potential for the inflaton%
\begin{equation}
V(\Phi )=(\frac{6-q}{3})\frac{16\pi }{{m_{pl}}^{2}}\rho _{b\Phi }^{0}e^{(-%
\sqrt{\frac{q}{2}}(\Phi -\Phi _{0}))},  \label{1}
\end{equation}%
it is possible to generate an intergalactic magnetic field whose present
strength (depending on values of parameters of the model) lies between $%
10^{-65}$ to $10^{-10}$ on a scale of $1/1000$ that of the Hubble scale. In (%
\ref{1}) $m_{Pl}$ is the planck mass, $\Phi ^{0}$ is the value of the scalar
field and $\rho _{b\Phi }^{0}$ is the homogeneous scalar field energy
density when the scale factor is $a_{0}$. Thus in this scenario one can have
the desired galactic magnetic field by resorting to the dynamo mechanism.

Considering next possibility (ii), if we add gauge non-invariant terms to
the action, the $U(1)$ gauge invariance will be broken. To avoid the
phenomenological disasters this can cause one can endow the photon with a
mass squared of the order of $H^{2}$ (well below present limits of
detectability). It has been shown \cite{Turner1} that such a term can create
primeval fields with strength as large as $r\sim 10^{-8}$ .

Gauge invariant modifications to the action have much better theoretical
motivation. For example, all $RF^{2}$ terms can be obtained by calculating
the effective Lagrangian for QED in curved space-time to one loop order \cite%
{Drummond}. At early times, when $R^{1/2}\sim H\sim \rho
_{tot}^{1/2}/m_{pl}\gg 10^{-11}m_{pl}$, these terms govern the behavior of
the electromagnetic field. However at later times, when $R^{1/2}\ll
10^{-11}m_{pl}$, they are negligible compared to the standard $-\frac{1}{4}%
F^{2}$ term. In a power-law inflationary background the amplitude of
large-scale fields is not large enough to be astrophysically interesting %
\cite{Turner1}.

The third possibility has proven to be promising for gauge theories with
large groups and a greater number of bosons than fermions. In such theories,
it has been shown that this mechanism for breaking conformal invariance in
quantum electrodynamics can create a sufficient amount of primordial
magnetic field \cite{Dolgov}. However, in the simplest version of the grand
unified $SU(5)$ model with three generations of fermions the magnetic field
produced is below the requirement of the dynamo mechanism.

Phase transitions at different cosmological epochs (grand unification, the
electroweak transition \cite{Vachaspati} or the quark confinement epoch \cite%
{Quashnock}) have also been considered. However, since the generating
mechanisms are causal, the coherence of the created magnetic field cannot be
larger than the particle horizon at the time of the phase transition.
Because all the above transitions occurred very early in the universe's
history, the comoving size of the horizon is rather small. The best case is
the QCD transition, for which the horizon corresponds to $\sim $ 1 a.u.
Consequently the real magnetic fields generated lack sufficient coherence.

The MSSM flat directions, made up of gauge invariant combinations of squarks
and sleptons, acquire non-vanishing vacuum expectation values (vev) during
inflation. These flat directions endow the standard model gauge fields with
mass and break the conformal invariance. The quantum fluctuations of these
flat directions, in contrast to the their classical vevs, induce
fluctuations in the gauge degrees of the freedom that cannot be gauged away.
The gauge field fluctuations that are stretched outside the horizon during
inflation, provide us with a seed (hyper)magnetic field after they re-enter
the horizon. They give rise to $U(1)_{em}$ magnetic field with strength of $%
10^{-30}$ G, as required by the dynamo mechanism \cite{Enqvist}.

Here we consider an alternative mechanism that is based on a hypothesis of a
minimal fundamental length scale. Minimal length breaks conformal invariance
and so it might be expected that primordial magnetic fields can be produced
during inflation. One suggestion \cite{Kempf} for implementing minimal
length into the inflationary scenario in the context of trans-planckian
physics \cite{Martin} is based on the hypothesis of a generalized
uncertainty principle:
\begin{equation}
\Delta x\Delta p\geq \frac{1}{2}\left( 1+\beta ~(\Delta p)^{2}\right) ,
\label{mur}
\end{equation}%
where $\sqrt{\beta }$ is the ultraviolet cutoff on the order of the Planck
or string length. In this paper we employ this formalism to implement
minimal length into the action of electrodynamics. This translates into a UV
cutoff which, once implemented, has the sole effect of modifying the
evolution of the electromagnetic field. As we will demonstrate, the
formalism is not able to create a squeezing effect for the electromagnetic
field. Therefore the energy density of the electromagnetic field attenuates
adiabatically, $\rho _{B}\propto a^{-4}$.

However, it has recently been shown \cite{Ashoorioon} that terms in the
action that are total time derivatives are not invariant under the influence
of the minimal length hypothesis \cite{Kempf}. Consequently such terms
contribute to the equations of motion of the matter fields. We consider in
this paper an example of a total time derivative that, under the influence
of the UV cutoff, causes the photon to gain a large negative mass during
inflation. This effect goes to zero at the end of inflation and so such a
mass is undetectable today. We shall show that this approach is successful
in providing the dynamo mechanism with sufficient primordial seed magnetic
field. Even in absence of the dynamo mechanism, one can adjust a free
parameter in the action to account for the observed magnetic field of
galaxies today.

\section{Cutoff Breaking of Conformal Invariance}

The existence of a preferred minimal length breaks the conformal invariance
of the background geometry. Here we will examine the effect of this
conformal breaking on the evolution of electromagnetic fields.

We introduce a fundamental length (i.e. the presence of a cut-off) in the
inflationary scenario via generalization of the quantum mechanical
commutation relation \cite{Kempf} \
\begin{equation}
\lbrack \mathbf{X},\mathbf{P}]=i\longrightarrow \lbrack \mathbf{X},\mathbf{P}%
]=i\left( f\left( \beta \right) \mathbf{1}+g(\beta )\mathbf{P}^{i}\mathbf{P}%
^{j}\right)  \label{2}
\end{equation}%
where $f\left( \beta \right) ,g\left( \beta \right) $ are functions such
that $f\left( 0\right) =1$ and $g\left( 0\right) =0$; their actual form is
determined by other criteria that we shall discuss below. This
generalization significantly modifies transplanckian physics, whose effects
are then manifest in the CMBR. Here we employ the above formalism to find
the effect this cutoff has on the evolution of magnetic fields.

We begin with the action of electromagnetism in an expanding curved
background
\begin{equation}
S=-\int \frac{1}{4}\sqrt{-g}g^{\mu \nu }g^{\alpha \beta }F_{\mu \alpha
}F_{\nu \beta }d^{3}\mathbf{y}d\eta  \label{3}
\end{equation}%
where the $y^{i}$'s are comoving spatial coordinates related to
the proper ones by $x^{i}=a(\eta )y^{i}$ and $\eta $ is the
conformal time. Assuming that the background is flat Friedmann
Robertson Walker, with the metric
\begin{equation}
ds^{2}=\left\{
\begin{array}{ll}
-dt^{2}+a^{2}(t)\sum\limits_{i=1}^{i=3}{dy^{i}}^{2} &  \\
a^{2}(\eta )(-d\eta ^{2}+\sum\limits_{i=1}^{i=3}{dy^{i}}^{2}), &
\end{array}%
\right.  \label{4}
\end{equation}%
one can write down the action in the following form:
\begin{equation}
S=\frac{1}{4}\int [2F_{oi}F_{oi}-F_{ik}F_{ik}]d^{3}\mathbf{y}d\eta  \label{5}
\end{equation}%
Roman indices $i$ and $k$ run from 1 to 3 and repeated indices are summed
over. The disappearance of the scale factor $a(\eta )$ is a consequence of
the conformal invariance of electromagnetism. By imposing the radiation
gauge $A_{0}=\partial _{i}A_{i}=0$, the above action can be rewritten in the
following form
\begin{equation}
S=\frac{1}{2}\int [(\partial _{0}\mathbf{A})^{2}-(\mathbf{\nabla }\times
\mathbf{A})^{2}]d^{3}\mathbf{y}d\eta .  \label{6}
\end{equation}%
in terms of the electromagnetic potential $A_{i}$. This action is the
familiar electrodynamic action,
\begin{equation}
S=\frac{1}{2}\int (\mathbf{E}^{2}-\mathbf{B}^{2})d^{3}\mathbf{y}d\eta ,
\label{7}
\end{equation}%
written in radiation gauge.

The most general form of the modified commuation relation (\ref{2}) that
breaks Lorentz invariance (see also \cite{Mota}) while preserving the
translational and rotational symmetry takes the following form \cite%
{Kempf,KMM}
\begin{equation}
\lbrack \mathbf{X}^{i},\mathbf{P}^{j}]=i\left( \frac{2\beta p^{2}}{\sqrt{%
1+4\beta p^{2}}-1}\delta ^{ij}+2\beta \mathbf{P}^{i}\mathbf{P}^{j}\right)
\label{8}
\end{equation}%
to first order in the parameter $\beta $. Here $p$ is the physical momentum.
We still assume that $[\mathbf{X}^{i},\mathbf{X}^{j}]=[\mathbf{P}^{i},%
\mathbf{P}^{j}]=0$. To impose this modified commuation relation, we rewrite
the action using proper spatial coordinates $x=a(\eta )y$ in the form:
\begin{equation}
S=\frac{1}{2}\int \frac{d\eta d^{3}\mathbf{x}}{2a^{3}}\left\{ \left( \left[
\partial _{\eta }+\frac{a^{\prime }}{a}\partial _{x^{i}}x^{i}-\frac{%
3a^{\prime }}{a}\right] \mathbf{A}\right) ^{2}-a^{2}\left( \nabla \times
\mathbf{A}\right) ^{2}\right\}  \label{15}
\end{equation}%
We can identify $-i\partial _{x^{i}}$ as the momentum operator, $\mathbf{P}%
^{i}$ , and $x^{i}$ as the position operator, $\mathbf{X}^{i}$. We can cast
the action (\ref{35}) to a simpler form:
\begin{equation}
S=\int \frac{d\eta }{2a^{3}}\left\{ \big(\mathbf{A},B^{\dag }(\eta )B(\eta )%
\mathbf{A})+a^{2}(\mathbf{P}\times \mathbf{A}\big)^{2}\right\} ,  \label{16}
\end{equation}%
where we have consolidated $(\partial _{\eta }+i\frac{a^{\prime }}{a}%
\sum_{i=1}^{3}\mathbf{P}^{i}\mathbf{X}^{i}-3\frac{a^{\prime }}{a})$ into a
new operator $B(\eta )$. Since $\partial _{i}A^{i}=0$, it means that $(%
\mathbf{P}\times \mathbf{A})^{2}=\mathbf{P}^{2}\mathbf{A}^{2}$. A suitable
vectorial Hilbert space representation of the new commutation relation can
be defined by using auxiliary variables $\rho ^{l}$:
\begin{equation}
\mathbf{X}^{l}\mathbf{A}(\rho )=i\partial _{\rho ^{l}}\mathbf{A}(\rho )
\label{9}
\end{equation}%
\begin{equation}
\mathbf{P}^{l}\mathbf{A}(\rho )=\frac{\rho ^{l}}{1-\beta \rho ^{2}}\mathbf{A}%
(\rho )  \label{10}
\end{equation}%
\begin{equation}
(A_{i}(\rho ),A_{j}^{\prime }(\rho ))=\int_{\rho ^{2}<\beta ^{-1}}d^{3}\rho
A_{i}^{\ast }(\rho )A_{j}^{\prime }(\rho )  \label{11}
\end{equation}

Recall that the usual quantum mechanical commutation relation, $%
[X^{i},P^{j}]=i\delta ^{ij}$ is defined on a Hilbert space with the
following representation
\begin{equation}
\mathbf{X}^{l}\mathbf{A}(\rho )=x^{l}\mathbf{A}(x)  \label{12}
\end{equation}%
\begin{equation}
\mathbf{P}^{l}\mathbf{A}(\rho )=-i\partial _{x^{l}}\mathbf{A}(x)  \label{13}
\end{equation}%
\begin{equation}
(A_{i}(x),A_{j}^{\prime }(x))=\int d^{3}\mathbf{x}A_{i}^{\ast
}(x)A_{j}^{\prime }(x)  \label{14}
\end{equation}%
Ultimately the action takes the following form:
\begin{equation}
S=\int d\eta \int_{\rho ^{2}<\beta ^{-1}}d^{3}\rho \frac{1}{2a^{3}}\left\{
|(\partial _{\eta }-\frac{a^{\prime }}{a}\frac{\rho ^{i}}{1-\beta \rho ^{2}}%
\partial _{\rho ^{i}}-\frac{3a^{\prime }}{a})\mathbf{A}|^{2}-\frac{a^{2}\rho
^{2}|\mathbf{A}|^{2}}{(1-\beta \rho ^{2})^{2}}\right\}  \label{17}
\end{equation}

The presence of $\rho $ derivatives means that the $\rho $ modes are
coupled. However we can find new variables ($\tilde{\eta},\tilde{k}$)
\begin{eqnarray}
\tilde{\eta} &=&\eta ,  \notag  \label{18} \\
\tilde{k}^{i} &=&a\rho ^{i}e^{(-\beta \rho ^{2}/2)}
\end{eqnarray}%
where the $\tilde{k}$ modes decouple because
\begin{equation}
\partial _{\eta }-\frac{a^{\prime }}{a}\frac{\rho ^{i}}{1-\beta \rho ^{2}}%
\partial _{\rho ^{i}}=\partial _{\tilde{\eta}}  \label{19}
\end{equation}%
We will use the common index notation $\bar{A}_{\tilde{k}}$ for those
decoupling modes. The $\tilde{k}$ modes coincide with the usual comoving
modes on large scales, i.e., only for small $\rho ^{2}$. This means that the
comoving k modes that are obtained by scaling, $k^{i}=ap^{i}$, decouple at
large distances and couple at small distances. The action now takes the form
\begin{equation}
S=\int d\tilde{\eta}\int_{\tilde{k}<a^{2}/e\beta }d^{3}\tilde{k}\mathcal{L}
\label{20}
\end{equation}%
where%
\begin{equation*}
\mathcal{L=}\frac{1}{2}\nu \left\{ {\left| \left( \partial _{\eta }-3\frac{%
a^{\prime }}{a}\right) \mathbf{\bar{A}}_{\tilde{k}}\right|}^{2} -\mu \left|
\mathbf{\bar{A}}_{\tilde{k}}\right| ^{2}\right\}
\end{equation*}%
We have defined $\mu $ and $\nu $ as
\begin{equation}
\mu (\eta ,\rho )\equiv \frac{a^{2}\rho ^{2}}{(1-\beta \rho ^{2})^{2}}
\label{22}
\end{equation}%
\begin{equation}
\nu (\eta ,\rho )\equiv \frac{e^{3\beta \rho ^{2}/2}}{a^{6}(1-\beta \rho
^{2})}  \label{23}
\end{equation}%
It is convenient to express these functions in terms of the Lambert $W$
function (defined so that $W(x)e^{W(x)}=x$ \cite{Coreless})
\begin{equation}
\mu (\eta ,\tilde{k})=-\frac{a^{2}}{\beta }\frac{W(\zeta )}{(1+W(\zeta ))^{2}%
}  \label{24}
\end{equation}%
\begin{equation}
\nu (\eta ,\tilde{k})=\frac{e^{-3W(\zeta )/2}}{a^{6}(1+W(\zeta ))}
\label{25}
\end{equation}%
where $\zeta =-\beta \tilde{k}^{2}/a^{2}$. The equation of motion for the
action (\ref{20}) is:
\begin{equation}
\bar{\mathbf{A}}_{\tilde{k}}^{\prime \prime }+\frac{\nu ^{\prime }}{\nu }%
\bar{\mathbf{A}}_{\tilde{k}}^{\prime }+\bigg(\mu -3\frac{\nu ^{\prime }}{\nu }(\frac{%
a^{\prime }}{a})-3\big(\frac{a^{\prime }}{a}\big)^{\prime }-9(\frac{%
a^{\prime }}{a})^{2}\bigg)\bar{\mathbf{A}}_{\tilde{k}}=0  \label{26}
\end{equation}%
The operations of Fourier transforming and of scaling from proper position
coordinates do not commute \cite{Kempf}. Hence the field variable $\bar{%
\mathbf{A}}_{\tilde{k}}$ is different from that commonly employed
in the literature, $\mathbf{A}_{\tilde{k}}$ by a factor of
$a^{3}$:
\begin{equation}
\bar{\mathbf{A}}_{\tilde{k}}=a^{3}{\mathbf{A}}_{\tilde{k}}
\label{27}
\end{equation}%
Taking into account eq.(\ref{27}) and introducing a new variable $\theta
:=a^{6}\nu $, we obtain
\begin{equation}
\mathbf{A}_{\tilde{k}}^{\prime \prime }+\frac{\theta ^{\prime }}{\theta }%
\mathbf{A}_{\tilde{k}}^{\prime }+\mu \mathbf{A}_{\tilde{k}}=0;  \label{28}
\end{equation}%
as the equation of motion for scalar perturbations in presence of a minimal
length cutoff.

The solutions to equation (\ref{28}) are constrained by the Wronskian
condition which follows from the canonical commutation relation between $%
\mathbf{A}_{\tilde{k}}$ and its conjugate momentum, $\mathbf{\Pi }_{\tilde{k}%
}=\theta \mathbf{A}_{\tilde{k}}^{\prime }$
\begin{equation}
\lbrack A_{\tilde{k}}^{i},\Pi _{\tilde{k}^{\prime }}^{j}]=i\delta
^{ij}\delta ^{3}(\tilde{k}-\tilde{k}^{\prime })  \label{29}
\end{equation}%
or equivalently
\begin{equation}
A_{\tilde{k}}^{i}A_{\tilde{k}}^{^{\prime }j\ast }-A_{\tilde{k}}^{i\ast }A_{%
\tilde{k}}^{^{\prime }j}=i\theta ^{-1}\delta ^{ij}  \label{30}
\end{equation}%
During the de Sitter phase, $a=-1/H\eta $ and so $\zeta =-\beta H^{2}\tilde{k%
}^{2}\eta ^{2}$. $\sigma =\sqrt{\beta }H$ is the ratio of the cutoff to the
Hubble parameter during inflation. The factors $\mu $ and $\theta ^{\prime
}/\theta $ have the following expansions in the limit in which the mode is
outside the horizon ($\tilde{k}\eta \ll 1$):
\begin{eqnarray}
\mu (\eta ,\tilde{k}) &=&\tilde{k}^{2}+3\beta ^{2}H^{2}\tilde{k}^{4}\eta
^{2}+\cdots  \label{31} \\
\frac{\theta ^{\prime }(\eta ,\tilde{k})}{\theta (\eta ,\tilde{k})}
&=&5\beta H^{2}\tilde{k}^{2}\eta +12\beta ^{2}H^{4}\tilde{k}^{4}\eta
^{3}+\cdots
\end{eqnarray}%
In that regime, the modes satisfy the following equation:
\begin{equation}
\mathbf{A}_{\tilde{k}}^{\prime \prime }+\tilde{k}^{2}\mathbf{A}_{\tilde{k}}=0
\label{32}
\end{equation}%
Thus $\mathbf{A}_{\tilde{k}}\propto e^{i\tilde{k}\eta }$ and $\mathbf{B}%
_{i}=\epsilon _{ilm}F_{lm}/a^{2}$ where $F_{lm}=\partial _{l}A_{m}-\partial
_{m}A_{l}$. As a result $\rho _{B}$ varies like $a^{-4}$, which is the
adiabatic result.

Superficially this mechanism is unable to amplify the cosmic magnetic
fields. However it has been shown that this method of implementing the
cut-off in the action has an ambiguity: total time derivatives no longer
reduce to pure boundary terms \cite{Ashoorioon}. In continuous space-time,
the presence of such a boundary term does not affect the evolution of the
electromagnetic potential. However the operator $\partial _{\eta }$ that
acts on the electromagnetic potential inside the total time derivative
transforms to $B(\eta )$ in the proper spatial coordinates. Since the
modification of the commutation relation between $X^{i}$ and $P^{j}$ affects
how this operator acts upon $A_{\mu }$, this procedure of implementing
minimal length will not keep such total time derivatives invariant.

Fortunately another option is available. It can be shown that any boundary
term in physical space transforms to a non-boundary term in momentum space
in the following manner:
\begin{equation}
\int (f(a,A))^{\prime }d^{3}\mathbf{y}d\eta \rightarrow \int \theta (\eta ,%
\tilde{k})(f(a,A_{\tilde{k}}))^{\prime }d^{3}\tilde{k}d\eta  \label{bnd-trns}
\end{equation}%
and so it is possible that they may contribute to the equation of motion in
such a way that the above behavior of the magnetic field is modified. Since
we wish to break the conformal invariance, we endow the photon with a mass
term \cite{Woodard,Prokopec}. This implies that $f(a,A)=g(a,a^{\prime
},\ldots )A^{\mu }A_{\mu }$. However we also do not want to modify the
behavior of the photon in the well-understood part of the history of the
universe, namely the radiation and matter dominated eras. Since during these
eras the scale factor respectively behaves as $\eta $ and $\eta ^{2}$, we
assume that $g(a)\propto a^{\prime \prime \prime }$. So far the proposed
boundary term adds to Eq.(\ref{32}) a term $\propto -\frac{a^{\prime \prime
\prime }\theta ^{\prime }}{a^{2}\theta }\mathbf{A}_{\tilde{k}}$ which looks
like $\mathbf{A}_{\tilde{k}}/\eta $ as the mode crosses outside the horizon.
Recalling the equation of motion for scalar fluctuations, $u_{k}$ (see for
e.g. \cite{Ashoorioon})
\begin{equation}
u_{k}+(k^{2}-\frac{z^{\prime \prime }}{z})u_{k}=0, \label{scalar}
\end{equation}%
the term that creates the amplification is $z^{\prime \prime
}u_{k}/z $ which behaves like $u_{k}/\eta ^{2}$ as the mode is far
outside the horizon. Hence we multiply the previous term with
another factor of $a$ to produce the desired behavior.

Summarizing, the proposed boundary term is:
\begin{equation}
\triangle S=\frac{1}{M_{1}^{2}}\int (A^{\mu }A_{\mu }a^{\prime \prime \prime
}a)^{\prime }d^{3}\mathbf{y}d\eta   \label{33}
\end{equation}%
where $\mu $ runs over space-time indices $0\cdots 3$ and $M_{1}$ is an
arbitrary constant with dimensions of mass whose presence keeps $\triangle S$
dimensionless. We can write this covariantly as
\begin{equation*}
\triangle S=\frac{1}{M_{1}^{2}}\int \nabla _{\alpha }(\Xi \zeta ^{\alpha })%
\sqrt{g}d^{3}\mathbf{y}d\eta
\end{equation*}%
where $\Xi $ is the scalar function
\begin{equation*}
\Xi =\frac{1}{3}\left( A^{\mu }A_{\mu }\right) \left( \nabla ^{2}K-2\nabla
^{\mu }\nabla ^{\nu }K_{\mu \nu }\right) \sqrt{\zeta \cdot \zeta }
\end{equation*}%
where $K_{\mu \nu }$ is the extrinsic curvature of the boundary surface
whose normal is $n_{\mu }=\left( a(\eta ),\vec{0}\right) $ and $\zeta
^{\alpha }=\left( 1,\vec{0}\right) $ is the conformal Killing vector of the
spacetime. Also, one can express  $a^{\prime \prime \prime }$ in the
following form:
\begin{equation}
a^{\prime \prime \prime }=a^{4}H^{3}(1-2q+j)  \label{a'''}
\end{equation}%
where $q$ and $j$ are respectively the deceleration and jerk parameters
defined as \cite{visser}
\begin{eqnarray}
q &=&-\frac{\ddot{a}}{aH^{2}},  \label{deceleration} \\
j &=&\frac{\dddot{a}}{aH^{3}},  \label{jerk}
\end{eqnarray}%
where dot denotes differentiation with respect to the physical time. The
presence of such a term modifies the propagator of the photon only during
inflation. The vertices and propagator of the electron do not get modified
at any time. Therefore, the amplitude for the diagrams that describe photon
splitting \cite{photon-split}, $\gamma \rightarrow n\gamma $, remain intact
and hence abide with the current bounds that exist on photon splitting \cite%
{Bound}.

The equation of motion for $\mathbf{A}_{\tilde{k}}$ derived from the
variation of the cutoff-modified action $S+\triangle S$ is
\begin{equation}
\mathbf{A}_{\tilde{k}}^{\prime \prime }+(\tilde{k}^{2}-\frac{1}{M_{1}^{2}}%
\frac{a^{\prime \prime \prime }\theta ^{\prime }}{a\theta })\mathbf{A}_{%
\tilde{k}}=0,  \label{34}
\end{equation}%
During the de Sitter expansion, $a=-1/H\eta $. For modes outside the
horizon, $\tilde{k}\eta \ll 1$, and eq.(\ref{34}) reduces to
\begin{equation}
\mathbf{A}_{\tilde{k}}^{\prime \prime }-\frac{n}{\eta ^{2}}\mathbf{A}_{%
\tilde{k}}=0  \label{35}
\end{equation}%
where $n=30\sigma ^{2}\tilde{k}^{2}/M_{1}^{2}$. In this limit we have $|%
\mathbf{A}_{\tilde{k}}|\propto \eta ^{m_{\pm }}$ where $m_{\pm }=\frac{1}{2}%
(1\pm \sqrt{1+4n})$. Here $\sigma =\sqrt{\beta }/H^{-1}$, where $\sqrt{\beta
}$ is the minimal length associated with the ultraviolet cutoff and $H$ is
the Hubble constant during inflation. The fastest growing solution during
the de Sitter phase is proportional to $\eta ^{p}$ or equivalently $a^{-p}$,
where we set $p=m_{-}$ . Note that for $p=-1$ ($n=2$), $|\mathbf{A}_{\tilde{k%
}}|$ varies like $a$ and $\rho _{B}\propto a^{-2}$, which is the
superadiabatic result. The evolution of electromagnetic waves during
reheating and the matter-dominated (MD) era is described by the same
equation as (\ref{34}) with $a\propto \eta ^{2}$, whereas in the radiation
dominated (RD) epoch $a\propto \eta $. In these three epochs the effective
mass of the photon vanishes and $\rho _{B}\propto a^{-4}$. During RD, MD,
and reheating, the electromagnetic field behaves as it does in absence of
the cut-off.

The ratio of the energy stored in the $k$-th mode of quantum fluctuations, $%
\rho _{B}(k)$, to the total energy density of the universe, $\rho _{tot}$,
at first horizon-crossing, $a=a_{1}$, is approximately equal to $%
[M/m_{Pl}]^{4}$. Here $M^{4}$ is the vacuum energy density during inflation.
Such a quantum fluctuation will be excited during the de Sitter expansion,
ref.\cite{Gibbons}, and can be treated as a classical fluctuation in the
electromagnetic field when it crosses outside the horizon. After
horizon-crossing $\rho _{B}(k)$ varies as $a^{-2(p+2)}$ while the total
energy density of the universe remains constant, $\rho _{tot}\propto M^{4}$.
Since the extra term added to the equation of motion is zero during
reheating, in the MD and RD epochs the stored energy density in the $k$-th
mode magnetic fluctuation attenuates adiabatically, $\rho _{B}\propto a^{-4}$%
. In reheating and the MD era, the energy density of the universe decreases
as $a^{-3}$ whereas in the RD epoch the total energy density of the universe
diminishes as $a^{-4}$. Therefore the invariant ratio, $\rho _{B}(k)/\rho
_{\gamma }$ on the scale $\lambda $ is:
\begin{equation}
r\simeq e^{-2N(\lambda )(p+2)}\bigg[\frac{M}{m_{Pl}}\bigg]^{8/3}\bigg[\frac{%
T_{RH}}{m_{Pl}}\bigg]^{4/3},  \label{36}
\end{equation}%
where $N(\lambda )$ is the number of e-folds the universe expands between
the first horizon crossing of the comoving scale $\lambda $ and the end of
inflation. It is given by the following equation \cite{Turner3}:
\begin{equation}
N(\lambda )=45+\ln \lambda _{Mpc}+\frac{2}{3}\ln (M_{14})+\frac{1}{3}\ln
(T_{10})  \label{37}
\end{equation}%
and $M=M_{14}10^{14}$ GeV, $T_{RH}=T_{10}10^{10}$ GeV. Plugging
this equation back into Eq.(\ref{36}), one obtains
\begin{equation}
r\simeq (7\times 10^{25})^{-2(p+2)}\bigg[\frac{M}{m_{Pl}}\bigg]^{-4p/3}\bigg[%
\frac{T_{RH}}{m_{Pl}}\bigg]^{-2p/3}\lambda _{Mpc}^{-2(p+2)}  \label{38}
\end{equation}%
The above formula is correct regardless of whether horizon re-crossing takes
place at the RD or MD eras. Note that we have normalized our comoving scales
such that today physical scales are equal to comoving scales, i.e. $%
a_{today}=1$.

\begin{figure}[t]
\includegraphics[angle=270, scale=0.36]{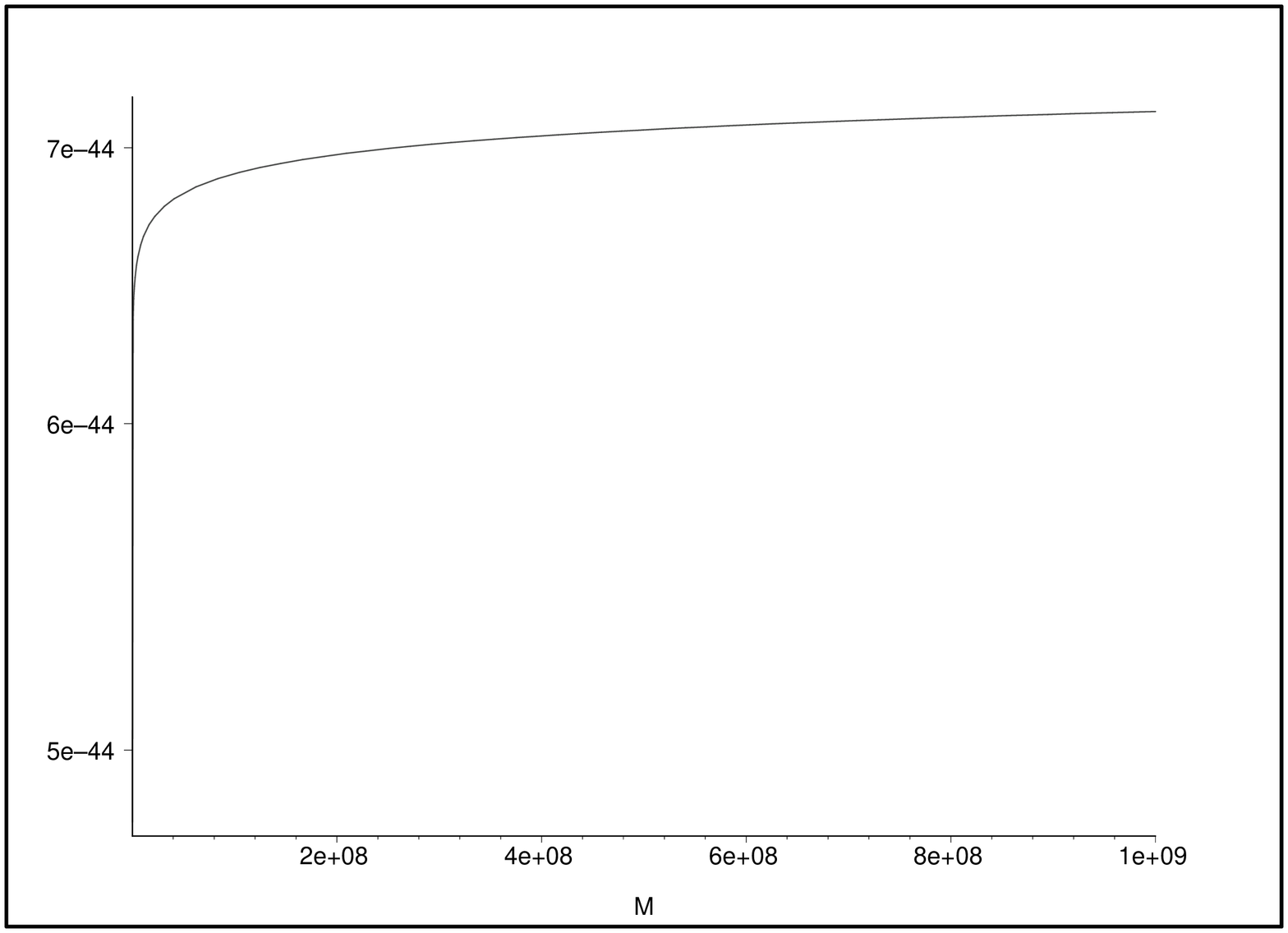}
\includegraphics[angle=270,
scale=0.36]{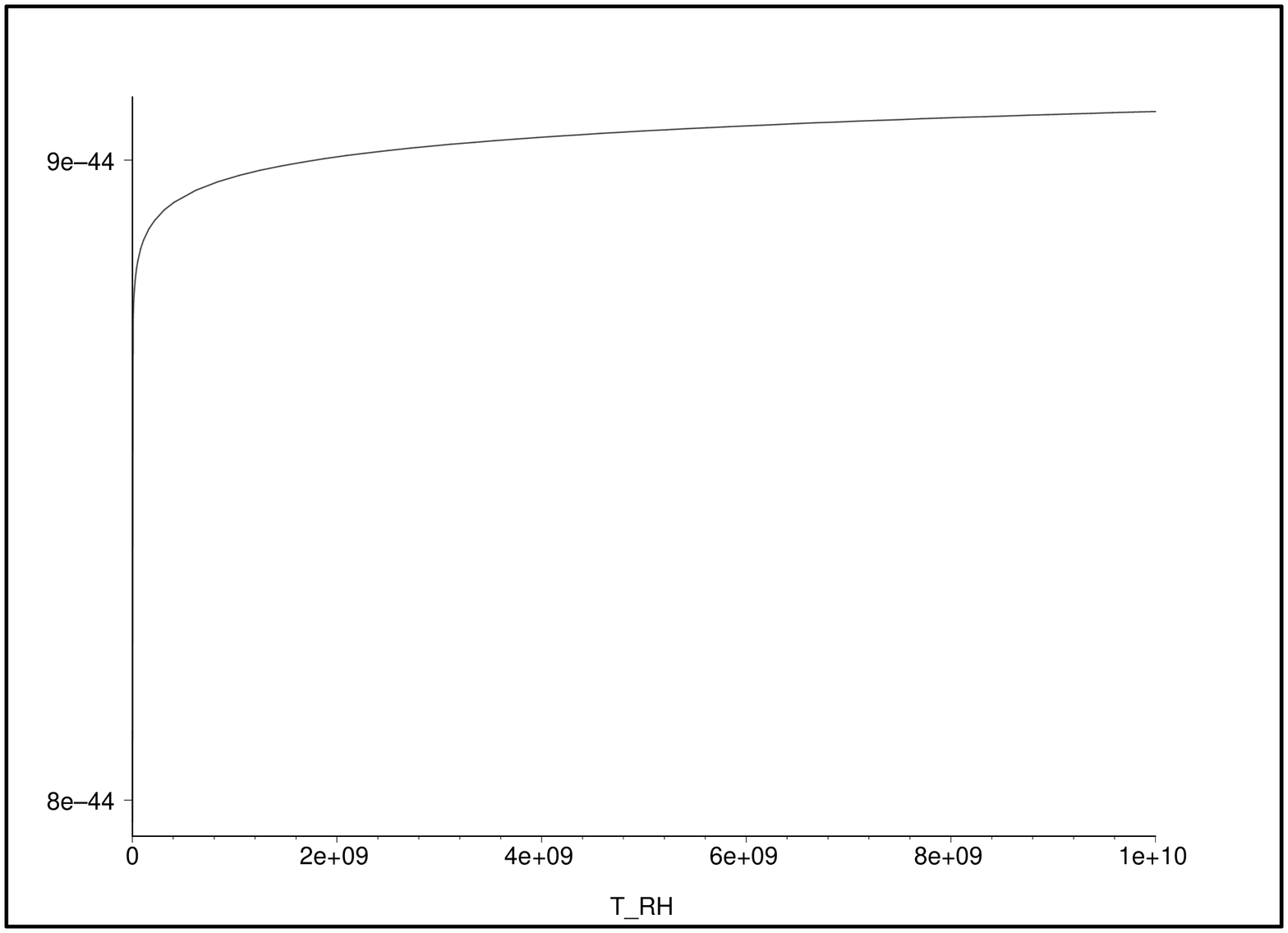}
\caption{The left figure shows the dependence of $M_{1}$ on $M$, the energy
scale of inflation. The right figure shows how $M_{1}$ varies as $T_{RH}$
changes. For all physically acceptable values of $M$ and $T_{RH}$, $%
M_{1}\sim 10^{-43}$GeV.}
\label{fig3}
\end{figure}

Two constraints on $M$ and $T_{RH}$ should hold in any viable scenario of
inflation. First, to prevent the production of long-wavelength gravitons
that distort the microwave background radiation beyond its upper limit of
anisotropy, $M<10^{-2}m_{Pl}$ . Second, $M$ and $T_{RH}$ should be greater
than $1$ GeV so that radiation domination takes place before nucleosynthesis.

To trigger the dynamo mechanism, there must be sufficient seed magnetic
field at cosmologically interesting scales. This condition could be used to
determine the value of $M_{1}$. Assuming that the required seed magnetic
field has been substantial on galaxy-scales, $\lambda \sim 1$Mpc, one can
obtain a relation between $M_{1}$ and other relevant parameters of the
problem:
\begin{equation}
M_{1}\simeq \frac{1.6\times 10^{-38}\big[\ln (T_{RH}/m_{Pl})+3\ln b+2\ln
(M/m_{Pl})\big]\sigma }{\bigg(\big[\ln r+4\ln b\big]\big[3\ln r+18\ln b+2\ln
(T_{RH}/m_{Pl})+4\ln (M/m_{Pl})\big]\bigg)^{1/2}}Gev  \label{39}
\end{equation}%
where $b=7\times 10^{-25}$. If $M,T_{RH}$ and $\sigma $ are specified, one
can obtain the corresponding values of $M_{1}$. In table 1, we have
tabulated the results for different values of $M,T_{RH}$ corresponding to
different scenarios of inflation and some values of $r$ required to initiate
astrophysically interesting phenomena. As Fig.(1) and (2) show, for a fixed
value of $\sigma $ and all physically relevant values of $M$ and $T_{RH}$, $%
M_{1}$ does not vary too much. For $\sigma \sim 10^{-5}$\cite{Ashoorioon} we
find $M_{1}\sim 10^{-43}$GeV. Since the coupling of the added term is
proportional to $M_{1}^{-2}$, the smallness of $M_{1}$ indicates that the
coupling of electromagnetic field to the curvature of the expanding
background, due to the existence of minimal length, has been enormous during
the inflationary era. However the coupling is extinguished in all other
epochs due to the special form of the interaction.

Although we should await a unified theory to determine how gravity is
coupled to the other fields of nature, this phenomenological scenario
suggests that the enigmatic primordial magnetic fields might have their
origin in the special characteristics of \ space-time at high energies ( See
also ref.\cite{Sheikh-jabbari} on how non-commutativity of the space-time
might help us account for primeval magnetic fields).
\begin{table}[tbp]
\begin{center}
\begin{tabular}{cccc}
\hline\hline
&  &  &  \\
$T_{RH}$(GeV) & $r|_{\lambda=1Mpc}$ & $M_{1}\times 10^{43}$(GeV) &  \\
\hline\hline
&  &  &  \\
$10^{17}$ & $10^{-8}$ & $0.5639$ &  \\
$10^{9}$ & $10^{-8}$ & $0.5103$ &  \\
$10^{17}$ & $10^{-34}$ & $0.7311$ &  \\
$10^{9}$ & $10^{-34}$ & $0.6638$ &  \\
$10^{17}$ & $10^{-56}$ & $0.9819$ &  \\
$10^{9}$ & $10^{-56}$ & $0.8973$ &  \\ \hline\hline
\end{tabular}%
\end{center}
\caption{Values for $M_{1}$ corresponding to different inflationary
scenarios and different values of $r=(\protect\rho _{B})/\protect\rho _{%
\protect\gamma }|_{1Mpc}$. Here $\protect\sigma $ is assumed to be
$10^{-5}$ and $M$ is held constant at $10^{17}$ GeV. $M_{1}$ does
not vary significantly for all interesting values of $M,T_{RH}$
and $r$. }
\end{table}

\section{Conclusion}

The origin of magnetic fields with $\mu $G strength that are observed on
intragalactic scales remains an intriguing mystery. As the observed magnetic
field is coherent on such cosmological scales, the first cosmological
process one might think of as being able to produce such prevalent fields is
inflation. However the conformal invariance of the electromagnetic field
prohibits the quantum fluctuations of the electromagnetic field from
squeezing and amplifying during inflation.

The existence of a minimal length breaks this conformal invariance. We have
proposed a scenario based on this observation that can provide the requisite
initial magnetic seed for the astrophysical dynamo mechanism. With a proper
choice of the free parameter within the theory one can avoid the need for
the dynamo mechanism.

The scenario is based on the observation that incorporating minimal length
at the level of first quantization, as was done for the first time in \cite%
{Kempf}, does not render total time derivatives invariant under the
influence of minimal length. Therefore one can have actions that are
equivalent at the continuous space-time level, but are distinct from one
another once the presence of minimal length is introduced. We added a
prototype for such a total time derivative term to the action of
electromagnetism that respects the behavior of the photon throughout the
history of the universe except for the inflationary era. During inflation
this term induces a huge mass for the photon. We found that to match this
model with observation we must tune the free parameter of the model, $M_{1}$%
, to be extremely small. Since $M_{1}^{-2}$ is proportional to the coupling
of electromagnetism to the background geometry during inflationary epoch,
the small size of $M_{1}$ is indicative of gravity and electromagnetism
being strongly coupled at that time. We note that the numerical value of $%
M_{1}$ is approximately the inverse Hubble length, but we have found no
deeper explanation for this coincidence at the level of the model presented
here.

Of course it is conceivable that other gauge bosons of the standard model
can inherit the same tachyonic instability that we have considered for the
photon. \ However all other gauge bosons are non-Abelian and so will
experience screening effects that we expect will tend to dampen out this
instability \cite{Biro}. \ A detailed calculation of this effect remains an
interesting subject for future study.

Of course the main drawback for this model is its arbitrariness in the
choice of total time derivative. It would be really interesting if one were
able to find candidates from existing models of fundamental physics. Our
main goal here was that of demonstrating that specific characteristics of
space-time at Planckian epochs can create observable phenomena in the
universe at much later cosmological times.

\section*{Acknowledgements}

We are thankful to R. Brandenberger for helpful discussions. This work was
supported by the Natural Sciences \& Engineering Research Council of Canada.

\end{document}